\documentstyle[amssymb,preprint,aps]{revtex}

\def\a{\alpha}

\def\y{\eta}

\newcommand{\be}{\begin{equation}}
\newcommand{\ee}{\end{equation}}
\newcommand{\bn}{\begin{eqnarray}}
\newcommand{\en}{\end{eqnarray}}

\newcommand{\ba}{\begin{array}}
\newcommand{\ea}{\end{array}}

\begin{document}
\draft
\title{{\Large Quantization of the Relativistic Particle}}
\author{S.P. Gavrilov\thanks{%
Universidade Federal de Sergipe, Brasil; on leave from Tomsk Pedagogical
University, Russia; present e-mail: gavrilov@ufs.br} and D.M. Gitman\thanks{%
e-mail: gitman@fma.if.usp.br}}
\address{Instituto de F\'{\i}sica, Universidade de S\~ao Paulo\\
P.O. Box 66318, 05315-970 S\~ao Paulo, SP, Brasil}
\address{Instituto de F\'{\i}sica, Universidade de S\~ao Paulo\\
P.O. 66318, 05315-970 S\~{a}o Paulo, SP, Brasil}
\date{\today}
\maketitle

\begin{abstract}
We revise the problem of the quantization of relativistic particle,
presenting a modified consistent canonical scheme, which allows one not only
to include arbitrary backgrounds in the consideration but to get in course
of the quantization a consistent relativistic quantum mechanics, which
reproduces literally the behavior of the one-particle sector of the
corresponding quantum field. At the same time this construction presents a
possible solution of the well-known old problem how to construct a
consistent quantum mechanics on the base of a relativistic wave equation.
\end{abstract}


Already for a long time there exists a definite interest in quantization of
classical and pseudoclassical models of relativistic particles (RP). This
problem meets such difficulties as zero-Hamiltonian phenomenon and time
definition problem. Consideration in arbitrary electromagnetic and
gravitational backgrounds creates additional difficulties. The usual aim of
the quantization is to arrive in a way to a corresponding relativistic wave
equation without any attempt to demonstrate that a consistent quantum
mechanics is constructed, since there is a common opinion that the
construction of a such a mechanics on the base of relativistic wave
equations is not possible due to existence of infinite number of negative
energy levels, and due to existence of negative vector norms (in scalar
case), and these difficulties may be only solved in QFT \cite{FolWo50}. One
of possible approach to the canonical quantization of RP models was
presented in \cite{GitTy90a} on the base of a special gauge, which fixes
reparametrization gauge freedom. However, the difficulties with inclusion of
arbitrary backgrounds were not overcome and the consistent quantum mechanics
was not constructed. It turns out that the whole scheme of quantization,
which was used in that papers and repeated then in numerous works, has to be
changed essentially to make it possible to solve the above problems and to
construct a quantum mechanics which is consistent to the same extent to
which a one-particle description is possible in the frame of the
corresponding QFT. One of the main point of the modification is related to a
principally new realization of the Hilbert space. At the same time this
construction gives a solution of the above mentioned old problem how to
construct a consistent quantum mechanics on the base of a relativistic wave
equation Below we present a demonstration for a spinless particle case. The
spinning particle case and all long technical details may be found by a
reader in \cite{GavGi00}.

We start with a reparametrization invariant action of a spinless
relativistic particle interacting with gravitational and electromagnetic
backgrounds, 
\begin{equation}
S=\int_{0}^{1}Ld\tau ,\;L=-m\sqrt{\dot{x}^{\mu }g_{\mu \nu }(x)\dot{x}^{\nu }%
}-q\dot{x}^{\mu }g_{\mu \nu }(x)A^{\nu }(x),\;\dot{x}^{\mu }=dx^{\mu }/d\tau
.  \label{1}
\end{equation}
We select a special gauge $g_{0i}=0$\ (then $g^{00}=g_{00}^{-1}>0,%
\;g^{ik}g_{kj}=\delta _{j}^{i}$) of the metric and define canonical momenta 
\begin{equation}
p_{\mu }=\frac{\partial L}{\partial \dot{x}^{\mu }}=-\frac{mg_{\mu \nu }\dot{%
x}^{\nu }}{\sqrt{\dot{x}^{2}}}-qA_{\mu }\;,\;\;A_{\mu }=g_{\mu \nu }A^{\nu }.
\label{3}
\end{equation}
The discrete variable $\zeta =\pm 1$ is important for our consideration, 
\begin{equation}
\zeta =-{\rm sign}\left[ p_{0}+qA_{0}\right] .  \label{5}
\end{equation}
It follows from (\ref{3}):$\;{\rm sign}(\dot{x}^{0})=\zeta ,$ and there is a
constraint $\Phi _{1}=p_{0}+qA_{0}+\zeta \omega =0.$ The total Hamiltonian $%
H^{(1)}$ we construct according to a standard procedure \cite{Dirac64}, 
\begin{equation}
\dot{\eta}=\{\eta ,H^{(1)}\},\;\Phi _{1}=0,\;\lambda >0,\;\;H^{(1)}=\zeta
\lambda \Phi _{1},\;\eta =(x^{\mu },p_{\mu }),\;(\lambda =|\dot{x}^{0}|).
\label{11}
\end{equation}
$\Phi _{1}=0$ is a first-class constraint. A possible gauge condition, which
fixes only $\lambda $, has the form \cite{GitTy90a}: 
\begin{equation}
\Phi _{2}=x^{0}-\zeta \tau =0~.  \label{15}
\end{equation}
We study equations of motion to clarify the meaning of $\zeta $ (below for
simplicity $g_{\mu \nu }=\eta _{\mu \nu }={\rm diag}(1,-1,\dots ,-1)$).
Interpreting $\zeta \tau =x^{0}$ as a physical time, $\zeta p_{i}=P_{i}$ as
a physical momentum, $dx^{j}/d(\zeta \tau )=dx^{j}/dx^{0}=v^{j}$ as a
physical three-velocity, ${\cal P}_{i}^{kin}=P_{i}+(\zeta q)A_{i}$ as the
kinetic momentum, we may see that (\ref{11}) in the gauge (\ref{15}) read: 
\[
\frac{d{\mbox{\boldmath${\cal P}$\unboldmath}}_{kin}}{dx^{0}}=(\zeta
q)\left\{ {\bf E}+[{\bf v},{\bf H}]\right\} ,\;{\bf v}=\frac{{%
\mbox{\boldmath${\cal P}$\unboldmath}}_{kin}}{\sqrt{m^{2}+{%
\mbox{\boldmath${\cal P}$\unboldmath}}_{kin}^{2}}},\;\frac{d\zeta }{dx^{0}}%
=0,\;\;\zeta =\pm 1, 
\]
$\mbox{\boldmath${\cal P}$\unboldmath}^{kin}=\left( {\cal P}%
_{i}^{kin}\right) $. Thus, the classical theory describes both particle and
antiparticles with charges $\zeta q$. One can prove that for independent
variables $\mbox{\boldmath$\y$\unboldmath}=(x^{k},p_{k},\zeta )$ equations
of motion are canonical with an effective Hamiltonian ${\cal H}_{eff}$ 
\begin{equation}
\dot{\mbox{\boldmath$\y$\unboldmath}}=\{\mbox{\boldmath$\y$\unboldmath},%
{\cal H}_{eff}\},\;{\cal H}_{eff}=\left[ \zeta qA_{0}(x)+\omega \right]
_{x^{0}=\zeta \tau }\;.  \label{24}
\end{equation}

Commutation relations for the operators $\hat{X}^{k},\hat{P}_{k},\hat{\zeta}$%
, which correspond to the variables $x^{k},p_{k},\zeta $, we define
according to their Poisson brackets, and we assume the operator $\hat{\zeta}$
to have the eigenvalues $\zeta =\pm 1$ by analogy with the classical theory.
Thus, nonzero commutators are: $[\hat{X}^{k},\hat{P}_{j}]=i\hbar \delta
_{j}^{k}\,$ and $\hat{\zeta}^{2}=1$. As a state space we select one $R$,
whose elements $\mbox{\boldmath$\Psi$\unboldmath}\in R$ are ${\bf x}$%
-dependent four-component columns (${\bf x}=x^{i}$) 
\begin{equation}
\mbox{\boldmath$\Psi$\unboldmath}=\left( 
\begin{array}{c}
\Psi _{+1}({\bf x}) \\ 
\Psi _{-1}({\bf x})
\end{array}
\right) ,\;\Psi _{\zeta }({\bf x})=\left( 
\begin{array}{c}
\chi _{\zeta }({\bf x}) \\ 
\varphi _{\zeta }({\bf x})
\end{array}
\right) ,\;\zeta =\pm 1\,.  \label{a2}
\end{equation}
\ The inner product in $R$ is defined as follows: 
\begin{eqnarray}
&&\left( \mbox{\boldmath$\Psi$\unboldmath},\mbox{\boldmath$\Psi$\unboldmath}%
^{\prime }\right) =\left( \Psi _{+1},\Psi _{+1}^{\prime }\right) +\left(
\Psi _{-1}^{\prime },\Psi _{-1}\right) ,  \nonumber \\
&&\left( \Psi ,\Psi ^{\prime }\right) =\int \overline{\Psi }({\bf x})\Psi
^{\prime }({\bf x})d{\bf x}=\int \left[ \chi ^{\ast }({\bf x})\varphi
^{\prime }({\bf x})+\varphi ^{\ast }({\bf x})\chi ^{\prime }({\bf x})\right]
d{\bf x},\;\overline{\Psi }=\Psi ^{+}\sigma _{1}.  \label{a5}
\end{eqnarray}
We seek basic operators in block-diagonal form,$\;\hat{\zeta}={\rm bdiag}%
\left( I,-I\right) ,\;\hat{X}^{k}=x^{k}{\bf I},\;\hat{P}_{k}=\hat{p}_{k}{\bf %
I},\;\;\hat{p}_{k}=-i\hbar \partial _{k},$ where $I$ and ${\bf I}$ are $%
2\times 2$ and $4\times 4$ unit matrices respectively. A quantum Hamiltonian 
$\hat{H}_{\tau },$ which defines the evolution in $\tau ,$ \ is constructing
using its classical analog ${\cal H}_{eff}$ , 
\begin{eqnarray}
\hat{H}_{\tau } &=&\hat{\zeta}q\hat{A}_{0}+\hat{\Omega},\;\hat{\Omega}={\rm %
bdiag}\left( \left. \hat{\omega}\right| _{x^{0}=\tau }\,,\left. \hat{\omega}%
\right| _{x^{0}=-\tau }\right) ,\;\hat{\omega}=\left( 
\begin{array}{cc}
0 & M \\ 
G & 0
\end{array}
\right) ,  \nonumber \\
M &=&-\left[ \hat{p}_{k}+qA_{k}\right] \sqrt{-g}g^{kj}\left[ \hat{p}%
_{j}+qA_{j}\right] +m^{2}\sqrt{-g},\;G=\frac{g_{00}}{\sqrt{-g}}.  \label{a11}
\end{eqnarray}
The operator $\hat{A}_{0}={\rm bdiag}\left( \left. A_{0}\right| _{x^{0}=\tau
}\,I,\;\left. A_{0}\right| _{x^{0}=-\tau }\,I\,\right) $ is related to the
classical quantity $\left. A_{0}\right| _{x^{0}=\zeta \tau }\,,$ and $\hat{%
\Omega}$ is related to the classical quantity $\left. \omega \right|
_{x^{0}=\zeta \tau }$. Indeed, $\hat{\Omega}^{2}={\rm bdiag}\left( \left.
MG\right| _{x^{0}=\tau }I,\;\;\left. GM\right| _{x^{0}=-\tau }I\right) $
corresponds (in classical limit) to square of the classical quantity $\left.
\omega \right| _{x^{0}=\zeta \tau }$. Quantum states evolute in time $\tau $
in accordance with the Schr\"{o}dinger equation $i\hbar \partial _{\tau }%
\mbox{\boldmath$\Psi$\unboldmath}(\tau )=\hat{H}_{\tau }\mbox{\boldmath$%
\Psi$\unboldmath}(\tau ),\;$where the columns $\Psi _{\zeta }(\tau ,{\bf x}%
),\;$and the functions $\varphi _{\zeta }(\tau ,{\bf x}),\,\chi _{\zeta
}(\tau ,{\bf x})$ from (\ref{a2}) depend now on $\tau \,.$ As before we
believe that $x^{0}=\zeta \tau $ may be treated as physical time and
reformulate the evolution in its terms. At the same time we pass to another
representation of state vectors. 
\begin{eqnarray}
&&\mbox{\boldmath$\Psi$\unboldmath}(x^{0})=\left( 
\begin{array}{c}
\Psi (x) \\ 
\Psi ^{c}(x)
\end{array}
\right) ,\;\Psi (x)=\left( 
\begin{array}{c}
\chi (x) \\ 
\varphi (x)
\end{array}
\right) ,\;\;\Psi ^{c}(x)=\left( 
\begin{array}{c}
\chi ^{c}(x) \\ 
\varphi ^{c}(x)
\end{array}
\right) ,  \nonumber \\
&&\Psi (x)=\Psi _{+1}(x^{0},{\bf x}),\;\Psi ^{c}(x)=\Psi _{-1}^{\ast
}(-x^{0},{\bf x}),\;\;x=\left( x^{0},{\bf x}\right) \;.  \label{a31}
\end{eqnarray}
The inner product of two states $\mbox{\boldmath$\Psi$\unboldmath}(x^{0})$
and $\mbox{\boldmath$\Psi$\unboldmath}^{\prime }(x^{0})$ in such a
representation takes the form 
\begin{equation}
\left( \mbox{\boldmath$\Psi$\unboldmath},\mbox{\boldmath$\Psi$\unboldmath}%
^{\prime }\right) =\left( \Psi ,\Psi ^{\prime }\right) +\left( \Psi ^{c},{%
\Psi ^{c}}^{\prime }\right) ,  \label{a32}
\end{equation}
where $\left( \Psi ,\Psi ^{\prime }\right) $ is given by (\ref{a5}). In this
representation the operators $\hat{\zeta}$ and $\hat{X}^{k}$ retain their
form, whereas the Schr\"{o}dinger equation changes 
\begin{eqnarray}
&&i\hbar \partial _{0}\mbox{\boldmath$\Psi$\unboldmath}(x^{0})=\hat{H}%
_{x^{0}}\mbox{\boldmath$\Psi$\unboldmath}(x^{0}),\;\hat{H}_{x^{0}}={\rm bdiag%
}\left( \hat{h}(x^{0}),\hat{h}^{c}(x^{0})\right) \,,  \nonumber \\
\; &&\hat{h}(x^{0})=qA_{0}I+\hat{\omega},\,\;\hat{h}^{c}(x^{0})=\left. \hat{h%
}(x^{0})\right| _{q\rightarrow -q}=-\left[ \sigma _{3}\hat{h}(x^{0})\sigma
_{3}\right] ^{\ast }.  \label{a34}
\end{eqnarray}

In accordance to our interpretation $\hat{\zeta}$ is charge sign operator.
Let $\mbox{\boldmath$\Psi$\unboldmath}_{\zeta }$ be states with a definite
charge $(\zeta q)$,$\;\ \hat{\zeta}\mbox{\boldmath$\Psi$\unboldmath}_{\zeta
}=\zeta \mbox{\boldmath$\Psi$\unboldmath}_{\zeta }$. It is easily to see
that states $\mbox{\boldmath$\Psi$\unboldmath}_{+1}$ with the charge $q$
have $\Psi ^{c}=0$. Then the equation (\ref{a34}) reads $i\hbar \partial
_{0}\Psi =\hat{h}(x^{0})\Psi \,.$ In fact it is Klein-Gordon equation (KGE)
\ for the charge $q$ in first order form. It reproduces exactly the
covariant KGE for the scalar field $\varphi (x)$ with the charge $q$, 
\[
\left[ \frac{1}{\sqrt{-g}}\left( i\hbar \partial _{\mu }-qA_{\mu }\right) 
\sqrt{-g}g^{\mu \nu }\left( i\hbar \partial _{\nu }-qA_{\nu }\right) -m^{2}%
\right] \varphi =0\,,\;\left( \chi =\sqrt{-g}g^{00}\left( i\partial
_{0}-qA_{0}\right) \varphi \right) \,. 
\]
States $\mbox{\boldmath$\Psi$\unboldmath}_{-1}$ with charge $-q$ have $\Psi
=0$. In this case the equation (\ref{a34}) reads $i\hbar \partial _{0}\Psi
^{c}=\hat{h}^{c}(x^{0})\Psi ^{c},$ with the Hamiltonian $\hat{h}^{c}(x^{0}),$
i.e. the KGE for the charge $-q.$ The inner product (\ref{a32}) between two
solutions with different charges is zero. For two solutions with charges $q$
it takes the form of KGE scalar product for the case of the charge $q$. For
two solutions with charges $-q$ the inner product (\ref{a32}) is expressed
via KGE scalar product for the case of the charge $-q$. The Schr\"{o}dinger
equation (\ref{a34}) is totally charge invariant$.$

The eigenvalue problems for the Hamiltonians $\hat{h}$ and $\hat{h}^{c}$ in
time independent external backgrounds 
\begin{eqnarray}
\hat{h}\psi _{\varkappa ,n} &=&\epsilon _{\varkappa ,n}\psi _{_{\varkappa
,n}}\;,\;\;\left( \psi _{\varkappa ,n},\psi _{\varkappa ^{\prime },n^{\prime
}}\right) =\varkappa \delta _{\varkappa ,\varkappa ^{\prime }}\delta
_{n,n^{\prime }}\,,\;\varkappa ,\varkappa ^{\prime }=\pm \,,  \nonumber \\
\hat{h}^{c}\psi _{\varkappa ,n}^{c} &=&\epsilon _{\varkappa ,n}^{c}\psi
_{\varkappa ,n}^{c}\,,\;\;\left( \psi _{k,n}^{c}\,,\psi _{\varkappa ^{\prime
},n^{\prime }}^{c}\right) =\varkappa \delta _{\varkappa ,\varkappa ^{\prime
}}\delta _{n,n^{\prime }},\;\psi _{\varkappa ,n}^{c}=-\sigma _{3}\psi
_{-\varkappa ,n}^{\ast }\,,\;\epsilon _{\varkappa ,n}^{c}=-\epsilon
_{-\varkappa ,n}\,,  \label{a33}
\end{eqnarray}
solve the eigenvalue problem of the Hamiltonian (\ref{a34}): 
\begin{eqnarray}
&&\hat{H}_{x^{0}}\mbox{\boldmath$\Psi$\unboldmath}=E\mbox{\boldmath$\Psi$%
\unboldmath}\,,\;\mbox{\boldmath$\Psi$\unboldmath}=\left( %
\mbox{\boldmath$\Psi$\unboldmath}_{\varkappa ,n}\,;\;\mbox{\boldmath$\Psi$%
\unboldmath}_{\varkappa ,n}^{c}\right) ,\;E=\left( \epsilon _{\varkappa
,n}\,;\,\epsilon _{\varkappa ,n}^{c}\right) \;\,,  \nonumber \\
&&\mbox{\boldmath$\Psi$\unboldmath}_{\varkappa ,n}=\left( 
\begin{array}{c}
\psi _{\varkappa ,n}({\bf x}) \\ 
0
\end{array}
\right) ,\;\mbox{\boldmath$\Psi$\unboldmath}_{\varkappa ,n}^{c}=\left( 
\begin{array}{c}
0 \\ 
\psi _{\varkappa ,n}^{c}({\bf x})
\end{array}
\right) ,\,\;\left( \mbox{\boldmath$\Psi$\unboldmath},\mbox{\boldmath$\Psi$%
\unboldmath}^{c}\right) =0  \nonumber \\
&&\left( \mbox{\boldmath$\Psi$\unboldmath}_{\varkappa ,n}^{c},%
\mbox{\boldmath$\Psi$\unboldmath}_{\varkappa ^{\prime },m}^{c}\right)
=\left( \mbox{\boldmath$\Psi$\unboldmath}_{\varkappa ,n},\mbox{\boldmath$%
\Psi$\unboldmath}_{\varkappa ^{\prime },m}\right) =\varkappa \delta
_{\varkappa \varkappa ^{\prime }}\delta _{nm\,}\,,\;\varkappa =\pm \;.
\label{d1}
\end{eqnarray}

On the Fig.1 we show typical spectra (one can keep in mind e.g. external
Coulomb field): 
\unitlength=0.8pt 
\begin{picture}(220,410)
\put(240,0){\line(0,1){410}}
\put(20,370){{\Large I}}
\put(20,30){{\Large II}}
\put(430,370){{\Large III}}
\put(430,30){{\Large IV}}
\put(0,200){\line(1,0){300}}
\put(110,0){\vector(0,1){400}}
\put(117,395){{\large $\varepsilon$}}
\multiput(105,265)(0,4){30}{\line(1,0){10}}
\multiput(105,135)(0,-4){30}{\line(1,0){10}}
\put(62,258){\phantom{$-$}$mc^2$}
\put(62,137){$-mc^2$}
{\thicklines \put(100,140){\line(1,0){20}}
             \put(100,225){\line(1,0){20}}
             \put(100,240){\line(1,0){20}}
             \put(100,249){\line(1,0){20}}
             \put(100,256){\line(1,0){20}}
             \put(100,260){\line(1,0){20}} }
\put(125,73) { $\left. {\rule{0pt}{56pt}} \right \}$ }
\put(125,302){ $\left. {\rule{0pt}{69pt}} \right \}$ }
\put(144,73){lower branch:} 
\put(144,55){$\varepsilon_{-,\a},\;\psi_{-,\a}$}
\put(144,302){upper branch:} 
\put(144,284){$\varepsilon_{+,n},\;\psi_{+,n}$}
\put(240,0){ 
 \begin{picture}(220,410)
 \put(0,200){\line(1,0){220}}
 \put(110,0){\vector(0,1){400}}
 \put(117,395){{\large $\varepsilon^{c}$}}
 \multiput(105,265)(0,4){30}{\line(1,0){10}}
 \multiput(105,135)(0,-4){30}{\line(1,0){10}}
 \put(62,258){\phantom{$-$}$mc^2$}
 \put(62,137){$-mc^2$}
 {\thicklines \put(100,140){\line(1,0){20}}
             \put(100,144){\line(1,0){20}}
             \put(100,151){\line(1,0){20}}
             \put(100,160){\line(1,0){20}}
             \put(100,175){\line(1,0){20}}
             \put(100,260){\line(1,0){20}} }
 \put(125,90)  { $\left. {\rule{0pt}{69pt}} \right \}$ }
 \put(125,320){ $\left. {\rule{0pt}{56pt}} \right \}$ }
 \put(144,90){lower branch:} 
 \put(144,72){$\varepsilon_{-,n}^{c},\; \psi^c_{-,n}$}
 \put(144,320){upper branch:} 
 \put(144,302){$\varepsilon_{+,\a}^{c},\; \psi^c_{+,\a}$}
 \end{picture}}

\end{picture}

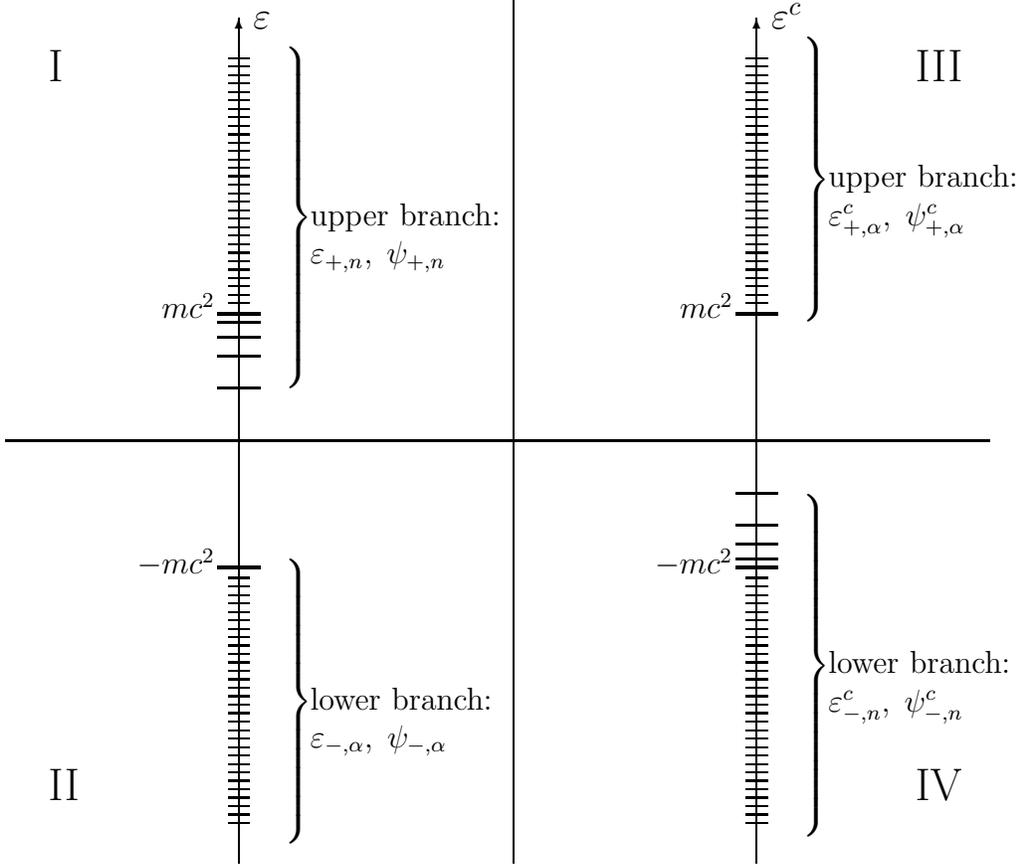
\noindent Fig.1.\ Energy spectra of KGE with charges $q$ and $-q$; I and II
- spectrum of $\hat{h}$, III and IV - spectrum of $\hat{h}^{c}$.

Let us compare results of the first quantization with one-particle sector of
the corresponding QFT. In course of second quantization the the classical
field (\ref{a31}) becomes operator $\hat{\Psi}(x)\;(\hat{\Psi}^{c}=-(\hat{%
\Psi}^{+}\sigma _{3})^{T},\;\hat{\overline{\Psi }}=\hat{\Psi}^{+}\sigma
_{1}),$%
\begin{equation}
\lbrack \hat{\Psi}(x),\hat{\overline{\Psi }}(y)]_{x^{0}=y^{0}}=\hbar \delta (%
{\bf x}-{\bf y})\;,\;i\hbar \partial _{0}\hat{\Psi}(x)=\hat{h}(x^{0})\hat{%
\Psi}(x)\,,\;i\hbar \partial _{0}\hat{\Psi}^{c}(x)=\hat{h}^{c}(x^{0})\hat{%
\Psi}^{c}(x).  \label{b1}
\end{equation}
In external backgrounds, which do not create particles from the vacuum, one
may define subspaces invariant under the evolution with definite numbers of
particles . Let us consider below only such backgrounds which do not depend
also on time to simplify the demonstration. A generalization to arbitrary
backgrounds, in which the vacuum remains stable, looks similar. One may
decompose the operator $\hat{\Psi}(x)$ in the complete set $\psi _{\varkappa
,n}$, then $\hat{\Psi}(x)=\sum_{n}\left[ a_{n}\psi _{+,n}(x)+b_{n}^{+}\psi
_{-,n}(x)\right] ,\;$ $[a_{n},a_{m}^{+}]=[b_{n},b_{m}^{+}]=\delta
_{nm},\;[a_{n},a_{m}]=[b_{n},b_{m}]=0\;.$ Thus, we get two sets of
annihilation and creation operators $a_{n},a_{n}^{+}$ and $b_{n},b_{n}^{+}$,
one of particles with a charge $q$ and another one of antiparticles with a
charge $-q$. Indeed, the Hamiltonian $\hat{H}^{QFT}$ of the QFT, charge
operator $\hat{Q}^{QFT}$ and particle number operator $\hat{N}$ read 
\begin{eqnarray*}
&&\hat{H}^{QFT}=\hat{H}_{R}^{QFT}+E_{0},\;\hat{H}_{R}^{QFT}=\sum_{n}\left[
\epsilon _{+,n}a_{n}^{+}a_{n}+\epsilon _{+,n}^{c}b_{n}^{+}b_{n}\right]
,E_{0}=-\sum_{n}\epsilon _{-,n}\,, \\
&&\hat{Q}^{QFT}=q\sum_{n}\left[ a_{n}^{+}a_{n}-b_{n}^{+}b_{n}\right] \;,\;%
\hat{N}=\sum_{n}\left[ a_{n}^{+}a_{n}+b_{n}^{+}b_{n}\right] \;,
\end{eqnarray*}
where $\hat{H}_{R}^{QFT}$ is a renormalized Hamiltonian. The Hilbert space $%
R^{QFT}$ of QFT is a Fock one. In the backgrounds under consideration each
subspace $R_{AB}^{QFT}$ of state vectors with the given number of particles $%
A$ and antiparticles $B$ is invariant under the time evolution. Now we are
in position to demonstrate that the one-particle sector of the QFT may be
formulated as a consistent relativistic quantum mechanics. We reduce the
space $R^{QFT}$ to a subspace of vectors which obey the condition $\hat{N}|%
\mbox{\boldmath$\Psi$\unboldmath}>=|\mbox{\boldmath$\Psi$\unboldmath}>$. It
is the subspace $R^{1}=R_{10}^{QFT}\oplus R_{01}^{QFT}.$ We call $R^{1}$
one-particle sector of QFT. All state vectors from the one-particle sector
have positive norms.\ The spectrum of the Hamiltonian $\hat{H}_{R}^{QFT}$ in
the space $R^{1}$ reproduces exactly one-particle energy spectrum of QFT (it
is situated on the areas ${\bf I}$ and ${\bf III}$ of the Fig.1.), 
\begin{equation}
\hat{H}_{R}^{QFT}|\mbox{\boldmath$\Psi$\unboldmath}>=E^{QFT}|%
\mbox{\boldmath$\Psi$\unboldmath}>\,,\;|\mbox{\boldmath$\Psi$\unboldmath}%
>=\left( a_{n}^{+}|0>;\;b_{n}^{+}|0>\right) \,,\;E^{QFT}=\left( \epsilon
_{+,n\,}\,;\;\epsilon _{+,n}^{c}\right) \,.  \label{c5}
\end{equation}
\ \ The dynamics of the one-particle sector may be formulated in a
coordinate representation, which is an analog of coordinate representation
in nonrelativistic quantum mechanics. Let us consider time-dependent states $%
|\mbox{\boldmath$\Psi$\unboldmath}(x^{0})>$ from the subspace $R^{1}$. One
may describe these states in the coordinate representation by four columns 
\begin{equation}
\mbox{\boldmath$\Psi$\unboldmath}(x^{0})=\left( 
\begin{array}{c}
\Psi (x) \\ 
\Psi ^{c}(x)
\end{array}
\right) ,\,\Psi (x)=<0|\hat{\Psi}(x)|\mbox{\boldmath$\Psi$\unboldmath}%
(0)>,\;\Psi ^{c}(x)=<0|\hat{\Psi}^{c}(x)|\mbox{\boldmath$\Psi$\unboldmath}%
(0)>,  \label{c10}
\end{equation}
where $\Psi (x)$ and $\Psi ^{c}(x)$\ have the form (\ref{a31}). The QFT
inner product reduces in this case to the inner product (\ref{a32}). One may
find expressions for the basic operators in the coordinate representation in
the one-particle sector. In particular, the Hamiltonian, the Schr\"{o}dinger
equation and charge operator $\hat{Q}^{QFT}$ are 
\begin{equation}
\hat{H}_{R}^{QFT}\rightarrow \hat{H}={\rm bdiag}\left( \hat{h}\,,\,\hat{h}%
^{c}\right) \;,\;i\hbar \partial _{0}\mbox{\boldmath$\Psi$\unboldmath}%
(x^{0})=\hat{H}\mbox{\boldmath$\Psi$\unboldmath}(x^{0}),\;\hat{Q}%
^{QFT}\rightarrow \hat{Q}=q\hat{\zeta}\,.  \label{c13}
\end{equation}
We meet, in fact, all the quantum mechanical constructions in the case under
consideration. The eigenvalue problem (\ref{c5}) in the coordinate
representation has the form (\ref{d1}), however the states $%
\mbox{\boldmath$\Psi$\unboldmath}_{-,n}$ and $\mbox{\boldmath$\Psi$%
\unboldmath}_{-,\alpha }^{c}$ are absent. It reproduces the spectrum (\ref
{c5})\ situated on the areas ${\bf I}$ and ${\bf III}$ of Fig.1. According
to superselection rules physical states are only those, which obey the
condition $\hat{Q}^{QFT}\mbox{\boldmath$\Psi$\unboldmath}_{\zeta }=\hat{\zeta%
}q\mbox{\boldmath$\Psi$\unboldmath}_{\zeta }=\zeta q\mbox{\boldmath$\Psi$%
\unboldmath}_{\zeta }\,,\;\;\zeta =\pm 1.$ This condition defines a physical
subspace $R_{ph}^{1}=R_{10}^{QFT}\cup R_{01}^{QFT}$ from the one-particle
sector $R^{1}$. Due to the structure of the operator $\hat{\zeta}$, the
states $\mbox{\boldmath$\Psi$\unboldmath}_{+1}$ contain only the upper half
of components, whereas, ones $\mbox{\boldmath$\Psi$\unboldmath}_{-1}$
contain only the lower half of components. One may see that the complete set 
$\mbox{\boldmath$\Psi$\unboldmath}_{+,n}$ and $\mbox{\boldmath$\Psi$%
\unboldmath}_{+,\alpha }^{c}$ consists only of physical vectors.

Returning to the first quantization, we may see that under certain
restrictions our quantum mechanics coincides literally with the one-particle
sector of QFT. These restrictions are related only to an appropriate
definition of the Hilbert space of the quantum mechanics. Indeed, all other
constructions in the quantum mechanics and in the one-particle sector of the
QFT in the coordinate representation coincide. Consider again the eigenvalue
problem (\ref{d1}) for the quantum mechanical Hamiltonian in the space $R$.
Its spectrum is wider than one of the QFT Hamiltonian in the space $R^{1}$.
To get the same spectrum as in QFT, we need to eliminate the vectors $%
\mbox{\boldmath$\Psi$\unboldmath}_{-,n}$ and $\mbox{\boldmath$\Psi$%
\unboldmath}_{-,\alpha }^{c}.$ We may define an analog of the space $R^{1}$
as a linear envelop of the vectors $\mbox{\boldmath$\Psi$\unboldmath}_{+,n}$
and $\mbox{\boldmath$\Psi$\unboldmath}_{+,\alpha }^{c}$ only. This space
does not contain negative norm vectors. The spectrum of the Hamiltonian (\ref
{a34})\ \ in such defined space coincides with one of the QFT Hamiltonian in
the one-particle sector. Reducing $R^{1}$ to $R_{ph}^{1}$, we get literal
coincidence between both theories. One may think that the reduction of the
space $R$ of the quantum mechanics to the space $R^{1}$ is necessary only in
the first quantization, thus an equivalence between the first and the second
quantization is not complete. However, the same procedure is present in the
second quantization. Indeed, besides the vectors (\ref{c5}) one could
consider hypothetically the following possibilities\ to form one-particle
states: 
\begin{eqnarray*}
&&1)\;a_{n}^{+}|0\rangle _{1},\;b_{n}|0\rangle
_{1},\;(a_{n}|0>_{1}=b_{n}^{+}|0>_{1}=0); \\
&&2)\;a_{n}|0\rangle _{2},\;b_{n}|0\rangle
_{2}\,,\;(a_{n}^{+}|0>_{2}=b_{n}^{+}|0>_{2}=0)\,; \\
&&3)\;a_{n}|0\rangle _{3},\;b_{n}^{+}|0\rangle
_{3},\;(a_{n}^{+}|0>_{3}=b_{n}|0>_{3}=0).
\end{eqnarray*}
$\;$ The states from the group $1)$ reproduce the usual spectrum of the KGE
which is situated on the areas ${\bf I}$ and ${\bf II}$ (see Fig.1). The
states from the group $2)$ reproduce the spectrum which is situated on the
areas ${\bf IV}$ and ${\bf II.}$ The states from the group $3)$ reproduce
the spectrum which is situated on the areas ${\bf IV}$ and ${\bf III}.$
These states are eliminated from the state space of the quantum field theory.

Thus, we see that the first quantization of classical actions of the
relativistic particle leads to relativistic quantum mechanics, which is
consistent to the same extent as corresponding quantum field theory in the
one-particle sector. Such quantum mechanics describes charged particles of
both signs (particles and antiparticles), and reproduces correctly their
energy spectra without infinite number of negative energy levels. No
negative vector norms need to be used in the corresponding Hilbert space.
There is also an important analogy with the second quantization. Both in
first and second quantizations we start with actions with a fixed charge and
in course of the quantizations we get charge symmetric theories where
particles and antiparticles are present on the same foot. It is also
important to stress that the first quantization and its comparison with
one-particle sector of the quantum field theory provides a very simple
solution for the well-known old problem: how to construct a consistent
quantum mechanics on the base of a relativistic wave equation? The solution
is very simple, instead to try to use the lower branch of the spectrum (area 
${\bf II}$ on Fig.1) one has to unite particle and antiparticle in one
multiplet on the base of Schr\"{o}dinger equation (\ref{a34}). Then the area 
${\bf III}$ appears naturally, and areas ${\bf II}$ and ${\bf IV}$ have to
be eliminated.

{\bf Acknowledgment} The authors are thankful to the foundations FAPESP,
CNPq (D.M.G), and FAPESE (S.P.G.) for support. Besides, S.P.G. thanks the
Department of Mathematical Physics of USP for hospitality, he thanks also
the Department of Physics of UEL (Brazil) for hospitality and the Brazilian
foundation CAPES for support on the initial stage of this work.


\begin{references}
\bibitem{FolWo50}  L. Foldy and S. Wouthuysen, Phys. Rep. {\bf 78} (1950)
29; H. Feshbach and F. Villars, Rev. Mod. Phys. {\bf 30} (1958) 24; S.
Weinberg, {\em The Quantum Theory of Fields} (Cambridge University Press,
New York 1995), Vol.~1.

\bibitem{GitTy90a}  D. Gitman and I. Tyutin, JETP Lett. {\bf 51} (1990) 214;
Classical Quantum Gravity {\bf 7} (1990) 2131.

\bibitem{GavGi00}  S.P. Gavrilov and D.M. Gitman, {\em Quantization of
Point-Like Particles and Consistent Relativistic Quantum Mechanics},
Publicacao IF/USP-1402/2000; hep-th/0003112.

\bibitem{Dirac64}  P. Dirac, {\em Lectures on Quantum Mechanics} (Yeshiva
University, New York 1964); D. Gitman and I. Tyutin, {\em Quantization of
Fields with Constraints} (Springer-Verlag, Berlin, 1990); M. Henneaux and C.
Teitelboim, {\em Quantization of Gauge Systems} (Princeton University Press,
Princeton 1992).
\end{references}
\end{document}